# Estimating marginal treatment effects from observational studies and indirect treatment comparisons: When are standardization-based methods preferable to those based on propensity score weighting?

**Harlan Campbell**

harlan.campbell@precisionvh.com

Department of Statistics, University of British Columbia, Vancouver, BC, Canada

Evidence Synthesis and Decision Modeling, PRECISIONheor, Vancouver, BC, Canada

**Julie E Park**

julie.park@precisionvh.com

Evidence Synthesis and Decision Modeling, PRECISIONheor, Vancouver, BC, Canada

**Jeroen P Jansen**

jeroen.jansen@precisionvh.com

School of Pharmacy, University of California, San Francisco, California, USA

Evidence Synthesis and Decision Modeling, PRECISIONheor, Vancouver, BC, Canada

**Shannon Cope**

shannon.cope@precisionvh.com

Evidence Synthesis and Decision Modeling, PRECISIONheor, Vancouver, BC, Canada

**Data availability statement** - The data that supports the findings of this study are available in the supplementary material of this article.

**Conflict of interest disclosure** - The authors have declared no conflict of interest.

**Funding statement** – The authors have no funding to declare.

# Estimating marginal treatment effects from observational studies and indirect treatment comparisons: When are standardization-based methods preferable to those based on propensity score weighting?

**Harlan Campbell**[*,1,2], **Julie E Park**[1], **Jeroen P Jansen**[1,3], **and Shannon Cope**[1]

[1] Evidence Synthesis and Decision Modeling, PRECISIONheor, Vancouver, BC, Canada
[2] Department of Statistics, University of British Columbia, Vancouver, BC, Canada
[3] School of Pharmacy, University of California, San Francisco, California, USA

In light of newly developed standardization methods, we evaluate, via simulation study, how propensity score weighting and standardization -based approaches compare for obtaining estimates of the marginal odds ratio and the marginal hazard ratio. Specifically, we consider how the two approaches compare in two different scenarios: (1) in a single observational study, and (2) in an anchored indirect treatment comparison (ITC) of randomized controlled trials. We present the material in such a way so that the matching-adjusted indirect comparison (MAIC) and the (novel) simulated treatment comparison (STC) methods in the ITC setting may be viewed as analogous to the propensity score weighting and standardization methods in the single observational study setting. Our results suggest that current recommendations for conducting ITCs can be improved and underscore the importance of adjusting for purely prognostic factors.

## 1 Introduction

Recently there has been much debate on the most appropriate target of inference when it comes to estimating relative treatment effects (Xiao et al., 2022; Remiro-Azócar, 2021; Liu et al. 2022). Some researchers believe that the most appropriate target is the marginal treatment effect while others make an argument in favour of the conditional treatment effect. The distinction between the two is important when the parameter of interest is non-collapsible, which is the case for the frequently used odds ratio (OR) and hazard ratio (HR). This is because with non-collapsible parameters, marginal and conditional estimands may not coincide due to non-linearity of the characteristic collapsibility function; see Greenland and Pearl (2011), Daniel et al. (2021) and Colnet et al. (2023). Even in the absence of confounding bias, with perfectly balanced treatment-groups, marginal and conditional estimates for the OR/HR will be different whenever variables included in the logistic/Cox outcome regression model are associated with the outcome (i.e., whenever covariates are "prognostic''); see Greenland et al. (1999). Put another way, in logistic and Cox proportional hazards regression models, conditioning on a prognostic variable inherently changes the relative treatment effect being estimated. In this article, we put aside the question of which is the most appropriate target of inference and focus on how best to obtain marginal estimates for the OR and HR.

The two most well-known approaches for obtaining covariate-adjusted marginal estimates are: (1) propensity score weighting (PSW) and (2) standardization (also known as g-computation); see Hernán and Robins (2006). Of course, in the absence of any confounders, the simplest way to obtain a marginal estimate is to simply fit a regression model without any covariates, thereby obtaining an unadjusted marginal estimate. However, the covariate-adjusted marginal estimate (obtained using either PSW or standardization) will be more efficient (i.e., have a smaller standard error/narrower confidence interval) than the unadjusted marginal estimate whenever the covariates in question are prognostic of the outcome; see Hernandez et al. (2004), Kahan et al. (2014), and Colantuoni and Rosenblum (2015).

---

[*]Corresponding author: e-mail: harlancampbell@gmail.com

PSW methods have origins that date back several decades (e.g., Horvitz and Thompson (1952) and Rosenbaum and Rubin (1983)) and they are ever increasing in popularity (Webster-Clark et al., 2021). Researchers have also been applying standardization-based methods for marginalizing adjusted ORs for several decades; see Zhang (2008) and Vansteelandt and Keiding (2011) who provide a brief history. However, for time-to-event outcomes, a satisfactory standardization procedure for marginalizing adjusted HRs has only recently been developed by Daniel et al. (2021).

Relative treatment effects can be estimated by means of a randomized controlled trial (RCT), an observational study, or by combining the findings of multiple RCTs, each comparing a subset of the treatment comparisons of interest. In the simplest form, this is known as an anchored indirect treatment comparison (ITC), but the general approach is known as network meta-analysis (NMA) (Tonin et al., 2017). It is often the case that individual participant data (IPD) is available for one of the RCTs included in an ITC, whereas data is restricted to published aggregate level data (AD) for the other trial(s). Two so called "IPD-AD methods" often used in ITCs of two trials are: (1) matching-adjusted indirect comparison (MAIC), a PSW-based approach (Signorovitch et al., 2010; Ishak et al., 2015); and (2) simulated treatment comparison (STC) (Ishak et al., 2015; Caro et al., 2010), a regression-based approach which (at least in theory) can be done with standardization in order to obtain a marginal estimate (Remiro-Azócar et al., 2021).

Marginal treatment effects are specific to a given target population (Remiro-Azócar et al., 2021) and often the population from which a sample is collected for a given study (i.e., the study population) differs from the target population (Josey et al., 2021; Dahabreh et al., 2020). For example, if interest is in determining the effectiveness of a policy expanding treatment to those who do not currently receive it, the study population in an observational study might include both those who do and do not receive the treatment while the target population would only include those who do not. In such a case, the average treatment effect in the untreated (ATU) would be of interest (Greifer and Stuart, 2021). In an ITC, the target population is typically defined (often for practical reasons) as the study population of the AD study.

Given the recent development of standardization-based methods for single studies, as well as for ITCs, it is important to compare their performance to PSW-based methods to determine the optimal approach in terms of bias and efficiency. Our goal in this article is to compare PSW-based and standardization-based approaches for obtaining estimates of the marginal OR and the marginal HR in (1) a single observational study where the target population is that of the untreated, and (2) in an anchored ITC of RCTs where the target population is that of the AD study. Note that the single observational study setting has been extensively studied in previous work (at least with regards to binary and continuous outcomes, e.g., Lunceford and Davidian (2004), Bang and Robins (2005), Delaney et al. (2017)), whereas the ITC setting is relatively less well understood. With this mind, to help better understand the ITC methods, we have organized the paper so that the MAIC and STC methods in the ITC setting may be viewed as analogous to the PSW and standardization methods in the single observational study setting. By laying out the material in this way, we hope to better understand the underlying principles at work and to better appreciate the notion that ITCs are "essentially observational findings across trials" (Higgins and Green, 2011).

This paper is organized as follows. In Section 2, we conduct a first simulation study to determine how PSW and standardization compare in a single observational study in the presence/absence of effect modifiers and different model misspecification. In Section 3, we conduct a simulation study to compare MAIC and STC approaches in an anchored ITC of RCTs. We summarize our overall findings and discuss implications and avenues for future research in Section 4.

## 2 Marginalization in a single comparative observational study

### 2.1 Objectives

Consider a single observational study for which IPD are available for the exposure, outcome, and covariates. If a covariate is a prognostic factor but not a confounder, adjusting for it will lead to more efficient estimation than excluding it and it is understood that PSW and standardization methods will typically result in identical gains in efficiency (Williamson, Forbes, and White, 2014; Daniel et al., 2021; Morris et al., 2022; Hernán and Robins, 2006). However, if a covariate is a confounder, standardization is thought to be a more efficient than PSW; see the simulation studies of Daniel et al. (2021) and Chatton et al. (2020), as well as earlier work by Bang and Robins (2005), and Tan (2007). The first objective of Simulation Study 1 is to shed further light on the relative efficiency of the standardization and PSW approaches.

Standardization will be unbiased when the outcome model is correctly specified and will be most efficient when the outcome model is parsimonious (i.e., when there is no overfitting). However, for PSW, defining a parsimonious propensity score model may not necessarily be advantageous. Schafer and Kang (2008), citing the theoretical work of Lunceford and Davidian (2004) and Rubin and Thomas (1996), explain that "[o]verfitting a propensity model can be beneficial, because propensities estimated by a rich model may carry more information about the potential outcomes in the sample than the true propensity scores, producing more efficient estimates." Most guidance suggests that any variables that are somehow related to the outcome should be included in the propensity score model, and all instrumental variables should be excluded; see Zhu et al. (2015), Austin et al. (2007) and Wooldridge (2016). The second objective of Simulation Study 1 is to investigate the consequences of model overfitting and model misspecification.

Daniel et al. (2021) present their standardization method, based on non-parametric Monte Carlo integration, for marginalizing adjusted HRs in the context of both a RCT and an observational study. However, Daniel et al. (2021) do not consider how their novel procedure might handle effect modifiers (i.e., treatment-covariate interaction effects). Chatton et al. (2020), in their simulation study, also fail to consider how using standardization to estimate the marginal OR may be impacted by effect modifiers. It also remains unclear how PSW performs in the presence of effect modifiers. Morris et al. (2022) note that PSW simply "does not handle" treatment-covariate interactions. However, a simulation study by Delaney et al. (2017) suggests that PSW does indeed provide unbiased estimates of the OR in the presence of (non-linear) treatment-covariate interactions. The third objective of Simulation Study 1 is therefore to investigate how PSW and standardization perform relative to one another in the presence of effect modifiers.

### 2.2 Data generation and mechanism

We simulate data as in Daniel et al. (2021) with certain modifications/additions. Let $X$ be the binary treatment indicator (equal to either 0 or 1), and $L$ be a measured continuous covariate.

For the binary outcome data, let $Y$ be the binary outcome (equal to either 0 or 1). We simulate $Y$ values from the following logistic regression model:

$$\text{logit}(\Pr(Y = 1|X = x, L = l)) = \alpha x + \beta_1 l + \beta_2 xl + \beta_3 xl^2,$$

where $\alpha$, $\beta_1$, $\beta_2$ and $\beta_3$ are regression coefficients.

As in Daniel et al. (2021), for TTE outcome data, we simulate individuals to enter the study uniformly over 2 years, and their event time then occurs at a random $Y$ years after their entry time. We simulate $Y$ values from a Weibull distribution with density function:

$$f(y|a, \sigma) = \left(\frac{a}{\sigma}\right)\left(\frac{y}{\sigma}\right)^{a-1} \exp\left(-\left(\frac{y}{\sigma}\right)^a\right),$$

where $a = 3/2$ and $\sigma = (0.1 \exp(\alpha x + \beta_1 l + \beta_2 xl + \beta_3 xl^2))^{-\frac{2}{3}}$, and $\alpha$, $\beta_1$, $\beta_2$, and $\beta_3$ are regression coefficients. All individuals having not experienced the event at 10 years since the start of the recruitment window are censored and the timescale for analysis is time since recruitment.

For both binary outcome and TTE outcome data, we simulate the covariate $L$ from a normal distribution with mean 0 and standard deviation of 1.5: $L \sim Normal(0, 1.5)$. Finally, we simulate the treatment exposure, $X$, from the following logistic model:

$$\Pr(X = 1|L = l) = \text{logit}^{-1}(\kappa_1 l + \kappa_2 l^2),$$

where $\kappa_1$ and $\kappa_2$ are regression coefficients. When $\kappa_1 = 1$ and $\kappa_2 = 0$, the mean of $L$ is unequal across treatment arms but the variance of $L$ is equal. Specifically, $L$ values amongst individuals in the treated arm will have mean of -0.80 and standard deviation of 1.27, and $L$ values amongst individuals in untreated arm will have mean of 0.80 and standard deviation of 1.27. When $\kappa_1 = 0$ and $\kappa_2 = 1$, the variance of $L$ is unequal across arms but the mean of $L$ is equal. Specifically, $L$ values amongst individuals in the treated arm will have mean of 0.00 and standard deviation of 0.71, and $L$ values amongst individuals in the untreated arm have mean of 0.00 and standard deviation 1.68.

For both the binary outcome data and the TTE outcome data, we consider eight different scenarios:

**SS -1A.** $L$ is an instrumental variable: $\beta_1 = 0$, $\beta_2 = 0$, $\beta_3 = 0$, $\kappa_1 = 1$, $\kappa_2 = 0$;

**SS -1B.** $L$ is a nonlinear instrumental variable: $\beta_1 = 0$, $\beta_2 = 0$, $\beta_3 = 0$, $\kappa_1 = 0$, $\kappa_2 = 1$.

**SS-2A.** $L$ is a confounder: $\beta_1 = 1$, $\beta_2 = 0$, $\beta_3 = 0$, $\kappa_1 = 1$, $\kappa_2 = 0$;

**SS-2B.** $L$ is a nonlinear confounder: $\beta_1 = 1$, $\beta_2 = 0$, $\beta_3 = 0$, $\kappa_1 = 0$, $\kappa_2 = 1$;

**SS-3A.** $L$ is a confounder and effect modifier: $\beta_1 = 1$, $\beta_2 = 1$, $\beta_3 = 0$, $\kappa_1 = 1$, $\kappa_2 = 0$;

**SS-3B.** $L$ is a nonlinear confounder, and effect modifier: $\beta_1 = 1$, $\beta_2 = 1$, $\beta_3 = 0$, $\kappa_1 = 0$, $\kappa_2 = 1$;

**SS-4A.** $L$ is a confounder and nonlinear effect modifier: $\beta_1 = 1$, $\beta_2 = 0$, $\beta_3 = 1$, $\kappa_1 = 1$, $\kappa_2 = 0$; and

**SS-4B.** $L$ is a nonlinear confounder and nonlinear effect modifier: $\beta_1 = 1$, $\beta_2 = 0$, $\beta_3 = 1$, $\kappa_1 = 0$, $\kappa_2 = 1$.

### 2.3 Estimands of interest

Unlike in Daniel et al. (2021), who target the "averaged treatment effect" (ATE), we target the average treatment effect in the untreated (ATU) (the effect that would be observed by switching every individual in the control arm to the treatment); see Greifer and Stuart (2021) for a formal definition. As target estimands, we consider the causal marginal log-odds ratio (log-OR) for the analysis of a binary outcome and the causal marginal log-hazard ratio (log-HR) for the analysis of a time-to-event outcome. We approximate the true marginal log-OR/log-HR values for each of the eight scenarios by simulating 10 million observations from the true sampling distributions. The log-HR values range from 0.079 to 1.807 and are listed alongside the results in Table 1. The log-OR values range from 0.066 to 2.360 and are listed alongside the results in Table A1.

Note that even when the conditional HR between an exposed individual and an unexposed individual with the same $L$ values is constant over time, the marginal HR will vary over time if $L$ is a prognostic factor (due to the non-collapsibility of the HR). As such, the definition of the marginal HR technically depends on the timeframe of interest. Daniel et al. (2021) explain this subtlety as follows: The marginal HR can be defined as "the probability limit (as the sample size $\to \infty$) of the marginal hazard ratio that would be estimated [with an unadjusted univariate Cox regression model] from an RCT of the chosen length". For the analysis in our simulation study, the timeframe of interest is 10 years (the point at which all remaining individuals are censored).

### 2.4 Methods to be compared

We compare five approaches: (A1) univariate (unadjusted) logistic/Cox regression model, (A2) PSW without a quadratic term, (A3) PSW with a quadratic term, (A4) standardization without an interaction term, and (A5) standardization with a $X$ by $L$ interaction term. Each approach specifies (either implicitly or explicitly) a given propensity score model and a given outcome model.

The **PSW method** involves two steps. First, a logistic regression propensity score model is fit with $X$, the binary treatment indicator, as the outcome and $L$ as the predictor. For A2, we have:

$$\text{logit}(\Pr(X = 1|L = l)) = \gamma_0 + \gamma_1 l, \tag{1}$$

and for A3, the model includes a quadratic term:

$$\text{logit}(\Pr(X = 1|L = l)) = \gamma_0 + \gamma_1 l + \gamma_2 l^2, \tag{2}$$

where $\gamma_0, \gamma_1$ and $\gamma_2$ are regression coefficients.

From this propensity score model, one estimates the so-called propensity scores, which are the probabilities of each subject receiving the treatment of interest conditional on their observed baseline covariate(s): $e_i = \Pr(X = 1|L = l_i)$. Note that, in the simulation study, the propensity score model without the quadratic term (equation (1)) will be misspecified whenever $\kappa_2 \neq 0$.

In the second step of PSW, one estimates the marginal log-OR (or log-HR) by fitting a univariate logistic (or Cox) regression outcome model with $Y$ as the outcome and $X$ as the only predictor and with individual weights based on the estimated propensity scores. Since we are targeting the ATU, the weights are calculated as: $w_{ATU} = X(1-e)/e + (1-X)$. Note that all "untreated" individuals receive a weight of 1. Treated individuals receive a weight inversely proportional to their odds of receiving treatment. See Austin and Stuart (2015) for further details on weighting. In the simulation study, this outcome model will be misspecified whenever $\beta_1 \neq 0$ or $\beta_2 \neq 0$ or $\beta_3 \neq 0$.

For the binary outcome data, **the standardization method** also involves two steps. The first step is fitting the outcome model, which in our case is a logistic regression conditional on $L$ (either with or without a treatment-covariate interaction term) to obtain estimates of the conditional probabilities: $\hat{\Pr}(Y = 1|X = 1, L = l_i)$ and $\hat{\Pr}(Y = 1|X = 0, L = l_i)$, for $i$ in 1,…, $N$, where $N$ is the number of observations in the study. In the simulation study, the outcome model without interaction term will be misspecified whenever $\beta_2 \neq 0$ or $\beta_3 \neq 0$ and the outcome model with interaction term will be misspecified whenever $\beta_3 \neq 0$. There is also an implicit propensity score model specified whereby treatment assignment is not dependent on any covariates (i.e., $X \sim 1$). In the simulation study, this propensity score model will be misspecified whenever $\kappa_1 \neq 0$ or $\kappa_2 \neq 0$.

In the second step of standardization, one averages over the empirical distribution of the covariate(s) within the untreated ($X=0$) group to obtain estimates of the marginal probabilities:

$$\hat{\Pr}_{ATU}(Y = 1|X = x) = \frac{1}{N} \sum_{i=1}^{N} \hat{\Pr}(Y = 1|X = x, L = l_i) \times I(X_i = 0),$$

for $x=0$ and $x=1$, where $I(X_i = 0)$ is the indicator function equal to 1 if $X_i = 0$, and zero otherwise. Finally, the covariate-adjusted estimator of the marginal log-OR is calculated as:

$$\log - OR = \log\left(\frac{\hat{\Pr}_{ATU}(Y = 1|X = 1)}{1 - \hat{\Pr}_{ATU}(Y = 1|X = 1)}\right) - \log\left(\frac{\hat{\Pr}_{ATU}(Y = 1|X = 0)}{1 - \hat{\Pr}_{ATU}(Y = 1|X = 0)}\right).$$

For TTE outcome data, the standardization procedure is similar. In the first step, the outcome model is a Cox regression model conditional on $L$ (either with or without a treatment-covariate interaction term, for A5 and A4 respectively). In the second step, to average over the empirical distribution of the covariate(s), one uses non-parametric Monte Carlo integration following the fourteen-step procedure proposed by Daniel et al. (2021). Briefly, the fourteen-step procedure works as follows. First, estimated marginal causal survival curves are used to simulate time-to-event data in a discrete manner for a hypothetical $X=1$ arm and a hypothetical $X=0$ arm with adjustments made to take into account the probability of censoring as well as the covariate distribution in the target population. This is repeated to obtain a total of $2m$ simulated survival times, where $m$ is a number much larger than the study sample size (the larger the value of $m$, the less Monte Carlo error there will be). Then, in the final steps, a univariate Cox model is fit via partial maximum likelihood estimation to the simulated data in order to obtain a covariate-adjusted estimator of the marginal log-HR. For details on how to implement each of the fourteen steps, see Section 4 of Daniel et al. (2021). We made one change to the steps as they are outlined by Daniel et al. (2021): we only proceed with simulating the censoring times if there are censored observations in both treatment arms. This small modification is necessary to avoid regular convergence failure.

### 2.5 Sample size, simulations, and performance measures

We set the study sample size to *N*=2000 and values for the covariate *L* are simulated anew in each simulation. Following the advice of Daniel et al. (2021), the standardization method is done with a value of *m* (the number of simulations used for Monte Carlo integration) set such that, upon repeating the analysis a second time with a value of *m* 10% larger, the point estimate of the marginal log-HR is within 0.009 of the original result (with a minimum of *m*=20000 and a maximum of *m*=100000).

For each of the 16 configurations (8 scenarios, 2 outcome types (binary and TTE)), we simulate 10000 datasets and calculate the marginal treatment effect estimate using each of the five approaches (A1-A5). Performance measures are defined by the sample mean and standard deviation of these 10000 estimates, which respectively reflect our simulation estimators of the mean and empirical standard error of each estimator. We also calculate the Monte Carlo SE of both simulation estimators using the formulae given in Morris et al. (2019). We quote our results to 3 decimal places.

### 2.6 Results

Results for the TTE outcome are listed in Table 1. Results for the binary outcome are all very similar and are listed in Table A1 in the Appendix.

With regards to bias, all five of the approaches appear to be unbiased if and only if *either* the propensity score model *or* the outcome model is correctly specified. For instance, in Scenario SS-2A in which *L* is a confounder, both the true outcome model (*Y~X+L*) and the true propensity score model (*X~L*) include a linear *L* term. The univariate unadjusted approach (A1) is biased in this scenario: the average estimate obtained with A1 is 2.081, substantially different than the target value of 0.768. This can be understood as bias due to unmeasured confounding. However, another way to think about this is that A1 is biased because neither the outcome model (*Y~X*), nor the propensity model (*Y~1*) that are defined by A1 include an *L* term and are therefore *both* misspecified. Despite having misspecified outcome models, both PSW methods (A2 and A3) define correctly specified propensity score models (*X~L* and *X~L+$L^2$*) and are therefore unbiased. Both standardization methods (A4 and A5) are also unbiased, despite having misspecified propensity score models, since the outcome models they define (*Y~X+L* and *Y~X+L+XL*) are correctly specified.

With regards to efficiency, when the univariate approach (A1) is unbiased (in SS-1A and SS-1B), it is at least as efficient as all the other methods. When both PSW and standardization approaches are unbiased, standardization is always at least as efficient as PSW. When both standardization approaches (A4 and A5) are unbiased (in SS-1A, SS-1B, SS-2A, and SS-2B), the standardization without interaction (A4) is always at least as efficient as the standardization with interaction (A5). This suggests that, as expected, there is a cost to overfitting the outcome model. However, overfitting the propensity score model may not always lead to lower efficiency. When both PSW approaches (A2 and A3) are unbiased (in SS-1A, SS-1B, SS-2A, SS-3A and SS-4A), the PSW with the quadratic term (A3) is more efficient than the PSW without the quadratic term (A2) in four out of the five scenarios (in SS-1A, SS-2A, SS-3A and SS-4A).

| Scenario | | True marginal logHR | Performance measure | A1. Univ. OM: $Y \sim X$ PS: $X \sim 1$ | A2. PSW OM: $Y \sim X$ PS: $X \sim L$ | A3. PSW OM: $Y \sim X$ PS: $X \sim L+L^2$ | A4. std. OM: $Y \sim X+L$ PS: $X \sim 1$ | A5. std. OM: $Y \sim X+L+XL$ PS: $X \sim 1$ |
|---|---|---|---|---|---|---|---|---|
| **SS-1A.** OM: $Y \sim X$ PS: $X \sim L$ | $\beta_1 = 0$ $\beta_2 = 0$ $\beta_3 = 0$ $\kappa_1 = 1$ $\kappa_2 = 0$ | 1.000 | Mean (MC error) | 1.001 (0.0005) | 1.003 (0.0008) | 1.003 (0.0008) | 1.001 (0.0006) | 1.001 (0.0006) |
| | | | Empirical SE (MC error) | 0.050 (0.0003) | 0.081 (0.0006) | 0.080 (0.0006) | 0.058 (0.0004) | 0.064 (0.0004) |
| **SS-1B.** OM: $Y \sim X$ PS: $X \sim L^2$ | $\beta_1 = 0$ $\beta_2 = 0$ $\beta_3 = 0$ $\kappa_1 = 0$ $\kappa_2 = 1$ | 1.000 | Mean (MC error) | 1.001 (0.0006) | 1.001 (0.0006) | 1.001 (0.0006) | 1.001 (0.0006) | 1.001 (0.0006) |
| | | | Empirical SE (MC error) | 0.057 (0.0004) | 0.058 (0.0004) | 0.065 (0.0005) | 0.058 (0.0004) | 0.058 (0.0004) |
| **SS-2A.** OM: $Y \sim X+L$ PS: $X \sim L$ | $\beta_1 = 1$ $\beta_2 = 0$ $\beta_3 = 0$ $\kappa_1 = 1$ $\kappa_2 = 0$ | 0.612 | Mean (MC error) | <span style="color:red">1.519 (0.0006)</span> | 0.617 (0.0012) | 0.621 (0.0010) | 0.612 (0.0004) | 0.612 (0.0005) |
| | | | Empirical SE (MC error) | <span style="color:red">0.057 (0.0004)</span> | 0.120 (0.0008) | 0.104 (0.0007) | 0.039 (0.0003) | 0.050 (0.0004) |
| **SS-2B.** OM: $Y \sim X+L$ PS: $X \sim L^2$ | $\beta_1 = 1$ $\beta_2 = 0$ $\beta_3 = 0$ $\kappa_1 = 0$ $\kappa_2 = 1$ | 0.759 | Mean (MC error) | <span style="color:red">0.349 (0.0005)</span> | <span style="color:red">0.362 (0.0004)</span> | 0.760 (0.0005) | 0.759 (0.0005) | 0.759 (0.0005) |
| | | | Empirical SE (MC error) | <span style="color:red">0.048 (0.0003)</span> | <span style="color:red">0.042 (0.0003)</span> | 0.052 (0.0004) | 0.046 (0.0003) | 0.048 (0.0003) |
| **SS-3A.** OM: $Y \sim X+L+XL$ PS: $X \sim L$ | $\beta_1 = 1$ $\beta_2 = 1$ $\beta_3 = 0$ $\kappa_1 = 1$ $\kappa_2 = 0$ | 0.079 | Empirical SE (MC error) | <span style="color:red">1.303 (0.0006)</span> | 0.083 (0.0010) | 0.086 (0.0008) | <span style="color:red">0.575 (0.0003)</span> | 0.080 (0.0004) |
| | | | Empirical SE (MC error) | 0.055 (0.0004) | 0.099 (0.0007) | 0.076 (0.0005) | <span style="color:red">0.034 (0.0002)</span> | 0.042 (0.0003) |
| **SS-3B.** OM: $Y \sim X+L+XL$ PS: $X \sim L^2$ | $\beta_1 = 1$ $\beta_2 = 1$ $\beta_3 = 0$ $\kappa_1 = 0$ $\kappa_2 = 1$ | 0.459 | Mean (MC error) | <span style="color:red">0.041 (0.0005)</span> | <span style="color:red">0.015 (0.0004)</span> | 0.460 (0.0005) | <span style="color:red">0.700 (0.0004)</span> | 0.460 (0.0005) |
| | | | Empirical SE (MC error) | <span style="color:red">0.046 (0.0003)</span> | <span style="color:red">0.040 (0.0003)</span> | 0.049 (0.0003) | <span style="color:red">0.041 (0.0003)</span> | 0.047 (0.0003) |
| **SS-4A.** OM: $Y \sim X+L+XL^2$ PS: $X \sim L$ | $\beta_1 = 1$ $\beta_2 = 0$ $\beta_3 = 1$ $\kappa_1 = 1$ $\kappa_2 = 0$ | 1.807 | Mean (MC error) | <span style="color:red">2.006 (0.0006)</span> | 1.810 (0.0007) | 1.808 (0.0007) | <span style="color:red">0.812 (0.0004)</span> | <span style="color:red">0.311 (0.0006)</span> |
| | | | Empirical SE (MC error) | 0.059 (0.0004) | 0.070 (0.0005) | 0.068 (0.0005) | <span style="color:red">0.041 (0.0003)</span> | <span style="color:red">0.065 (0.0005)</span> |
| **SS-4B.** OM: $Y \sim X+L+XL^2$ PS: $X \sim L^2$ | $\beta_1 = 1$ $\beta_2 = 0$ $\beta_3 = 1$ $\kappa_1 = 0$ $\kappa_2 = 1$ | 1.140 | Mean (MC error) | <span style="color:red">1.456 (0.0006)</span> | <span style="color:red">1.522 (0.0006)</span> | 1.141 (0.0006) | <span style="color:red">1.450 (0.0006)</span> | <span style="color:red">1.475 (0.0006)</span> |
| | | | Empirical SE (MC error) | <span style="color:red">0.058 (0.0004)</span> | <span style="color:red">0.058 (0.0004)</span> | 0.062 (0.0004) | <span style="color:red">0.056 (0.0004)</span> | <span style="color:red">0.056 (0.0004)</span> |

Rows SS-1A through SS-4B are grouped under **Observational Study**.

**Table 1** – Results from Simulation Study 1 for TTE outcome data. Values listed under "True marginal log-HR" were calculated by approximation, by simulating 10 million observations from the true Weibull sampling distributions. Numbers in red correspond to results for which the difference between the mean and the true marginal log-HR is more than 0.1.

## 3 Marginalization in an indirect treatment comparison

### 3.1 Objectives

Consider a new treatment, treatment *C*, which must be compared to an already established treatment, treatment *B*. Suppose treatment *B* has been evaluated in a RCT against a comparator, treatment *A* (e.g., standard of care or placebo) and only aggregate data (AD) from this RCT are available. Furthermore, suppose treatment *C* has also been evaluated against treatment *A* in a RCT and individual level data (IPD) are available; see lower portion ("Simulation Study 2") of Figure 1.

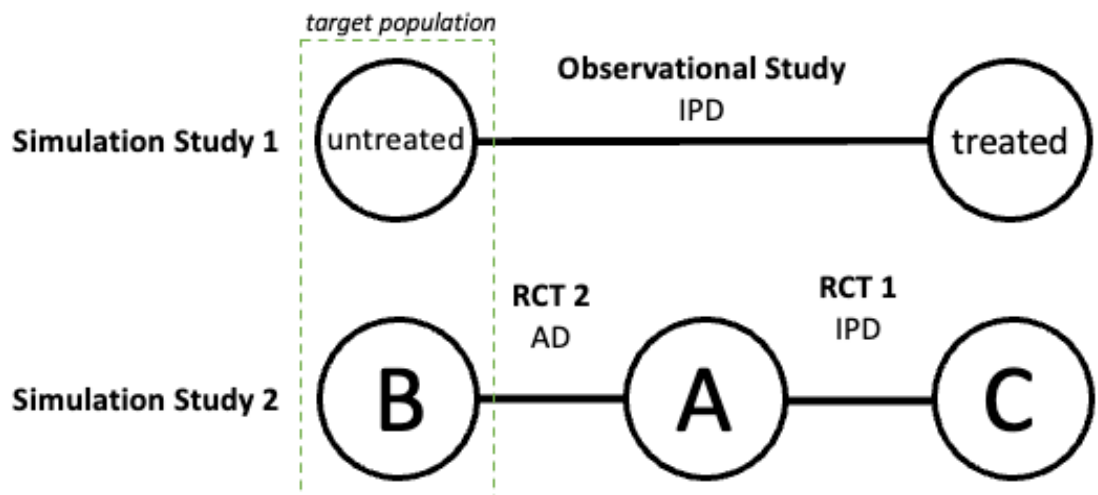

**Figure 1** – In Simulation Study 1, we consider a single observational study which compares a group of treated individuals and a group of untreated individuals. The interest is in determining the average treatment effect in the untreated (ATU). In Simulation Study 2, we consider an anchored ITC of two RCTs. Study 1 is an RCT comparing treatment A to treatment C for which individual participant-level data (IPD) are available. Study 2 is an RCT comparing treatment A to treatment B for which only aggregate-level data (AD) are available. The interest is in determining the average treatment effect in the Study 2 population.

Since treatments *B* and *C* have not been evaluated against each other directly within a single study, an anchored indirect treatment comparison (ITC) using IPD-AD methods must be performed to estimate the relative *B* vs. *C* treatment effect. Complicating matters further, the population in study 1 (the *A* vs. *C* study) may differ from the population in study 2 (the *A* vs. *B* study) such that the *B* vs. *C* treatment effect that would be observed in the study 1 population is not the same as the *B* vs. *C* treatment effect that would be observed in the study 2 population. Suppose the target population is that from the study 2 population as would be typical in an ITC. (Note that since study 2 is an RCT, the target population can equivalently be thought of as that of the *B* arm as is indicated in Figure 1.)

Recently, several simulation studies have attempted to compare the STC and MAIC approaches. Remiro-Azócar et al. (2021a) compared MAIC and STC for time-to-event outcomes using a simulation and noted that conventional STC is inappropriate for estimating the marginal effect since conventional STC targets the conditional effect and will therefore have "systematic bias as a result of the non-collapsibility of the log-hazard ratio." They identified the need for an "alternative formulation to STC that estimates a marginal treatment effect [...]", which was later accomplished in Remiro-Azócar (2022) who propose using g-computation (standardization) "as an extension to the conventional STC" and found that it provided "more precise and more accurate estimates than MAIC" for binary outcomes. For time-to-event outcomes, the g-computation extended STC approach proposed by Remiro-Azócar (2022) is less than ideal since it is limited to marginalization at a specific time-point. On this point, Remiro-Azócar (2022) speculate that the standardization methods of Daniel et al. (2021) may prove useful. As the first objective of Simulation Study 2, we follow through with this idea and consider the standardization method of Daniel et al. (2021) as an extension to conventional STC for the estimation of marginal HRs in the context of an ITC of RCTs for time-to-event outcomes. Note that this approach (using Daniel et al. (2021)'s standardization as an extension to the conventional STC) is rather novel and has yet to be applied in any published ITCs.

The distribution of effect modifiers may differ across studies in an ITC such that, unless one can properly adjust for these differences, estimation will be rendered invalid; see Jansen and Naci (2013). Therefore, understanding how best to adjust for effect modifiers to obtain unbiased marginal estimates is particularly important. The second objective of Simulation Study 2 is therefore to examine how the STC and MAIC techniques compare in terms of obtaining unbiased marginal OR/HR estimates in the presence of effect modifiers.

As in Simulation Study 1, we will also investigate the consequences of model misspecification. Recently, Vo (2023) demonstrated that the presence of a nonlinear effect modifier will cause the STC method (with a misspecified outcome model) to be "substantially biased," and that "MAIC does not perform much better." However, Vo (2023) only considered MAIC with matching on $1^{st}$ moments. As in Delaney et al. (2017), Vo (2023) specifically considered a treatment by covariate-squared interaction term (i.e., a $XL^2$ interaction term). Vo (2023)'s results would appear to perhaps contradict the conclusions of Signorovitch et al. (2010) who write that an "advantage of the proposed matching-adjusted [MAIC] estimate is its robustness to model misspecification." Clarifying matters with respect to the "robustness" of MAIC and STC is the third objective of Simulation Study 2.

The MAIC and STC methods are relatively new and as a result there is little guidance on best practices apart from the recommendations available is a National Institute for Health and Care Excellence Decision Support Unit (NICE DSU) Technical Support Document (Phillippo et al., 2016). Notably, Phillippo et al. (2016) recommend that when using STC for an anchored ITC, one should adjust for all imbalanced effect modifiers and that "further effect modifiers and prognostic variables may be adjusted for if this improves model fit to reduce standard error." When using MAIC for an anchored ITC, Phillippo et al. (2016) recommend adjusting for all effect modifiers but not adjusting for any purely prognostic variables ("all effect modifiers should be adjusted for to ensure balance and reduce bias, but no purely prognostic variables to avoid inflating standard error due to over-matching.") However, this recommendation seems to contradict the recent conclusions of Riley et al. (2022) who note that, for non-collapsible effect measures, failing to adjust for purely prognostic variables can lead to inconsistency when the distribution of prognostic factors differs across studies. Considering this seemingly conflicting advice, the fourth objective of Simulation Study 2 is to examine the consequences of failing to adjust for purely prognostic variables.

When using MAIC, there is also the question of whether it is best to implement the matching on only first moments, or on both first and second moments. According to Hatswell et al. (2020), who considered MAIC in the context of unanchored ITCs, matching on both first and second moments cannot be recommended ("in no scenarios did it provide a meaningful advantage, while also showing the potential for large errors"). However, others disagree with this recommendation. For instance, Petto et al. (2019) recommend that, if "different variances of an important effect modifier are observed", one should match on both the first and second moments, even though this will likely lead to wider confidence intervals. Most recently, Phillippo et al. (2021) summarize the current (lack of) understanding writing that: "the question of when, if at all, it is preferable or necessary to match higher moments between populations for MAIC remains an interesting area for further theoretical research and simulation studies." Answering this question will be the fifth objective of Simulation Study 2.

To summarize, the objectives of Simulation Study 2 are:

(1) Determine if using the standardization method of Daniel et al. (2021) as an extension to conventional STC for the estimation of marginal HRs in an anchored ITC of RCTs is effective.
(2) Determine how STC and MAIC techniques compare in terms of obtaining unbiased marginal OR/HR estimates in the presence/absence of unbalanced effect modifiers.
(3) Determine the consequences of model misspecification.
(4) Determine the consequences of failing to adjust for purely prognostic variables.
(5) Determine when, if at all, it is preferable or necessary to match on both the first and second moments (as opposed to only the first moment) when using MAIC.

### 3.2 Data generation and mechanism

As in Simulation Study 1, we consider both binary outcomes and TTE outcomes. We simulate individual participant level data (IPD) from "study 1" to mimic data from a RCT evaluating treatment *C* versus treatment *A*, and simulate aggregate level data from "study 2" to mimic data from a RCT evaluating treatment *B* versus treatment *A*.

We deliberately simulate data so that the scenarios in Simulation Study 2 correspond closely to the scenarios considered in Simulation Study 1. Specifically, in each scenario, the distribution of *Y*, *X*, and *L* in arm C of Study 1 (the "AC" IPD RCT) is the very same as the distribution of *Y*, *X*, and *L* in the treated arm of the single observational study considered Simulation Study 1. Similarly, the distribution of *Y*, *X*, and *L* in arm B of Study 2 (the "AB" AD RCT) is the very same as the distribution of *Y*, *X*, and *L* in the untreated arm of the single observational study. See Figure 1.

Let *X* be the binary treatment indicator (equal to either 0 (treatment *A*) or 1 (treatment *C* in study 1, and treatment *B* in study 2)), and let *L* be a measured continuous covariate. In both studies, we assume the treatment exposure, *X*, to be a balanced binary variable and therefore simulate $X \sim Bernoulli(0.5)$. The covariate *L* values are simulated according to the following procedure.

First, 100,000 values of *L* are simulated from a Normal distribution with mean 0 and standard deviation of 1.5. Then, for *i* in 1,…,100,000, the *i*-th *L* value, $l_i$, is randomly assigned to either an individual in the study 1 population or to an individual in the study 2 population with probabilities equal to:

$$\Pr(i \text{ in study } 1|l_i) = 1 - \Pr(i \text{ in study } 2|l_i), \text{ and}$$

$$\Pr(i \text{ in study } 2|l_i) = \text{logit}^{-1}(\kappa_1 l_i + \kappa_2 l_i^2),$$

where $\kappa_1$, and $\kappa_2$ are regression coefficients. Samples are then drawn from the study 1 and study 2 populations. Note that this procedure for simulating covariate values, while perhaps unusual, will allow us to consider scenarios for which the trial allocation model defined by the MAIC procedure is either correctly or incorrectly specified. When $\kappa_1 = 1$ and $\kappa_2 = 0$, the mean of $L$ is unequal across studies but the variance of $L$ is equal. Specifically, $L$ values amongst individuals in Study 1 will have mean of -0.80 and standard deviation of 1.27, and $L$ values amongst individuals in Study 2 will have mean of 0.80 and standard deviation of 1.27. When $\kappa_1 = 0$ and $\kappa_2 = 1$, the variance of $L$ is unequal across studies but the mean of $L$ is equal. Specifically, $L$ values amongst individuals in Study 1 will have mean of 0.00 and standard deviation of 0.71, and $L$ values amongst individuals in Study 2 have mean of 0.00 and standard deviation 1.68.

For binary outcome data, $Y$ is the binary outcome, equal to either 0 or 1. For study 1, $Y$ is simulated from the following logistic regression model:

$$\text{logit}(\Pr(Y = 1 | X = x, L = l)) = \alpha_{AC} x + \beta_1 l + \beta_{2AC} xl + \beta_{3AC} xl^2;$$

and for study 2, $Y$ is simulated from:

$$\text{logit}(\Pr(Y = 1 | X = x, L = l)) = \alpha_{AB} x + \beta_1 l + \beta_{2AB} xl + \beta_{3AB} xl^2,$$

where $\alpha_{AC}, \beta_1, \beta_{2AC}, \beta_{3AC}, \alpha_{AB}, \beta_{2AB}$, and $\beta_{3AB}$ are regression coefficients.

For TTE outcome data, individuals (in both studies) are simulated (as in Simulation Study 1) to enter the study uniformly over 2 years, and their event times are then simulated to occur at a random time $Y$ years after this entry time. The $Y$ times are then simulated from a Weibull distribution with density function:

$$f(y|a, \sigma) = \left(\frac{a}{\sigma}\right)\left(\frac{y}{\sigma}\right)^{a-1} \exp\left(-\left(\frac{y}{\sigma}\right)^a\right),$$

where $a = 3/2$, and, for study 1: $\sigma = (0.1 \exp(\alpha_{AC} x + \beta_1 l + \beta_{2AC} xl + \beta_{3AC} xl^2))^{-\frac{2}{3}}$; and for study 2: $\sigma = (0.1 \exp(\alpha_{AB} x + \beta_1 l + \beta_{2AB} xl + \beta_{3AB} xl^2))^{-\frac{2}{3}}$; where $\alpha_{AC}, \beta_1, \beta_{2AC}, \beta_{3AC}, \alpha_{AB}, \beta_{2AB}$, and $\beta_{3AB}$ are regression coefficients.

The aggregate data from study 2 consists only of the estimated log-OR/log-HR, $\hat{\theta}_{AB|Study2}$, and its standard error, $\text{SE}(\hat{\theta}_{AB|Study2})$, as well as the mean and standard deviation of the covariate $L$, $\hat{\mu}_{L2}$ and $\hat{\sigma}_{L2}$, respectively. Note that this information is typically available in a study's primary results and table of baseline characteristics.

We consider eight scenarios for both binary and TTE outcomes. In all eight scenarios, we set $\alpha_{AC} = 1$, $\alpha_{AB} = \beta_{2AB} = \beta_{3AB} = 0$, and each scenario is defined as follows:

**ITC-1A.** $L$ is entirely unrelated to $Y$, such that: $\beta_1 = 0, \beta_{2AC} = 0, \beta_{3AC} = 0$; the means of the covariate $L$ differ across studies and the variances are equal, with: $\kappa_1 = 1, \kappa_2 = 0$.

**ITC-1B.** $L$ is entirely unrelated to $Y$, such that: $\beta_1 = 0, \beta_{2AC} = 0, \beta_{3AC} = 0$; the means of the covariate $L$ are equal across studies and the variances differ, with: $\kappa_1 = 0, \kappa_2 = 1$.

**ITC-2A.** *L* is a prognostic factor, but not an effect modifier, such that: $\beta_1 = 1, \beta_{2AC} = 0, \beta_{3AC} = 0$; the means of the covariate *L* differ across studies and the variances are equal, with: $\kappa_1 = 1, \kappa_2 = 0$.

**ITC-2B.** *L* is a prognostic factor, but not an effect modifier, such that: $\beta_1 = 1, \beta_{2AC} = 0, \beta_{3AC} = 0$; the means of the covariate *L* are equal across studies and the variances differ, with: $\kappa_1 = 0, \kappa_2 = 1$.

**ITC-3A.** *L* is a prognostic factor, and an effect modifier, such that: $\beta_1 = 1, \beta_{2AC} = 1, \beta_{3AC} = 0$; the means of the covariate *L* differ across studies and the variances are equal, with: $\kappa_1 = 1, \kappa_2 = 0$.

**ITC-3B.** *L* is a prognostic factor, and an effect modifier, such that: $\beta_1 = 1, \beta_{2AC} = 1, \beta_{3AC} = 0$; the means of the covariate *L* are equal across studies and the variances differ, with: $\kappa_1 = 0, \kappa_2 = 1$.

**ITC-4A.** *L* is a prognostic factor, and a nonlinear effect modifier, such that $\beta_1 = 1, \beta_{2AC} = 0, \beta_{3AC} = 1$; the means of the covariate *L* differ across studies and the variances are equal, with: $\kappa_1 = 1, \kappa_2 = 0$.

**ITC-4B.** *L* is a prognostic factor, and a nonlinear effect modifier, such that $\beta_1 = 1, \beta_{2AC} = 0, \beta_{3AC} = 1$; the means of the covariate *L* are equal across studies and the variances differ, with: $\kappa_1 = 0, \kappa_2 = 1$.

### 3.3  Estimands of interest

The target estimand in Simulation Study 2 is the causal marginal *C* vs. *B* treatment effect that would be observed in the study 2 population, $\theta_{CB|Study2}$. We approximate the true $\theta_{CB|Study2}$ values for each of the scenarios by simulating 10 million observations from the true sampling distributions. The log-HR values range from 0.079 to 1.807 and are listed alongside the result in Table 2. The log-OR values range from 0.066 to 2.360 and are listed alongside the results in Table A2.

### 3.4  Methods to be compared

We compare five different estimation approaches for both the binary outcome and the TTE outcome data. Much like the approaches considered in Section 2 for a single observational study, each ITC approach specifies (either implicitly or explicitly) a given propensity score model for trial allocation, and a given outcome model. The five approaches are:

A1. Bucher method;

A2. MAIC (matching on 1$^{st}$ moment);

A3. MAIC (matching on 1$^{st}$ and 2$^{nd}$ moments);

A4. STC with standardization *without* a *X* by *L* interaction term; and

A5. STC with standardization *with* a *X* by *L* interaction term.

The **Bucher method** (Bucher et al., 1997) involves first fitting a unadjusted (logistic or Cox) regression outcome model (i.e., *Y~X*) to the study 1 IPD to obtain $\hat{\theta}_{AC|Study1,\text{univ}}$, the estimated marginal treatment effect (i.e., the log-OR or log-HR) of treatment *A* vs. *C* (in the study 1 population). This approach assumes that this is also a reasonable estimate for the marginal treatment effect of *A* vs. *C* in the study 2 population. With $\hat{\theta}_{AB|Study2}$ available from the study 2 AD, to obtain the estimated marginal treatment effect of *C* vs. *B* (in the study 2 population), we simply calculate: $\hat{\theta}_{CB|Study2,univ} = \hat{\theta}_{AC|Study1,univ} - \hat{\theta}_{AB|Study2}$. There is also an implicit propensity score model specified whereby trial selection is not dependent on any covariates (i.e., *Study* ~ 1).

The **MAIC** approach implicitly fits a propensity score model for trial allocation that is analogous to the propensity score model for treatment allocation used for PSW in the single study setting (Signorovitch et al., 2010; Cheng et al., 2020). While PSW (as discussed in Section 2) defined the propensity model for treatment assignment, MAIC defines a model for trial assignment. Amongst all the individuals in Study 1 and Study 2, suppose the probability that the *i*-th individual is included in Study 2 is related to their value of *L* according to the following logistic model:

$$\text{logit}(\Pr(i \text{ in } Study\ 2|l_i)) = \zeta_0 + \zeta_1 l_i,$$

where $\zeta_0$ and $\zeta_1$ are regression coefficients. Then, in order to obtain an estimate of $\theta_{AC|Study2}$, one could fit a weighted univariate (logistic or Cox) outcome model (with intercept term and covariate for treatment indicator) to the Study 1 IPD with weights defined by the estimated odds of inclusion in Study 2:

$$w_i = [\Pr(i \text{ in } Study\ 2|\ l_i)/(1 - \Pr(i \text{ in } Study\ 2|\ l_i)))] = \exp\left(\hat{\zeta}_0 + \left(\hat{\zeta}_1\ l_i\right)\right),$$

where $\hat{\zeta}_0$ and $\hat{\zeta}_1$ are estimates for $\zeta_0$ and $\zeta_1$ in the above logistic trial assignment model. Because IPD are unavailable for Study 2, $\hat{\zeta}_0$ and $\hat{\zeta}_1$ are obtained by the method of moments which involves solving a convex optimization problem that directly enforces balance in selected covariate moments. Let $\hat{A}$ be the value of A that minimizes the $g1(A)$ function, defined as:

$$g1(A) = \sum_{i=1}^{N_1} \exp\left(\left(l_i - \hat{\mu}_{L2}\right) \times A\right),$$

for *i* in 1,…,$N_l$. Then $\hat{\zeta}_0 = -\hat{\mu}_{L2}\hat{A}$ and $\hat{\zeta}_1 = \hat{A}$, and the weights are be defined as:

$$w_i = \exp\left(\hat{\zeta}_0 + \hat{\zeta}_1\ l_i\right) = \exp\left(\left(l_i - \hat{\mu}_{L2}\right) \times \hat{A}\right),$$

for *i* in 1,…,$N_l$; where $\hat{\mu}_{L2}$ is the sample mean of *L* in the study 2 sample. Using these weights corresponds to MAIC with "matching on the 1st moment". The MAIC can also be implemented with "matching on 1st and 2nd moments" in which one implicitly defines a trial assignment model with quadratic effects:

$$\text{logit}(\Pr(i \text{ in } Study\ 2|l_i)) = \zeta_0 + \zeta_1 l_i + \zeta_2 l_i^2,$$

where $\zeta_0$, $\zeta_1$ and $\zeta_2$ are regression coefficients. Estimates for these, $\hat{\zeta}_0$, $\hat{\zeta}_1$, and $\hat{\zeta}_2$, can be obtained with the method of moments as follows. Let $\hat{A1}$ and $\hat{A2}$ be the values of A1 and A2 that minimize the function, $g2(A1, A2)$, defined as:

$$g2(A1, A2) = \sum_{i=1}^{N_1} \exp((l_i - \hat{\mu}_{L2}) \times A1 + \left(l_i^2 - \hat{\mu}_{L2}^2 - \hat{\sigma}_{L2}^2\right) \times A2),$$

for $i$ in $1,\ldots,N_l$; where $\hat{\mu}_{L2}$ and $\hat{\sigma}_{L2}$ are the sample mean and sample standard deviation, respectively, of $L$ in the Study 2 sample. Recall that $\hat{\mu}_{L2}$ and $\hat{\sigma}_{L2}$ are available from the study 2 AD. Weights are then calculated as:

$$w_i = \exp\left(\hat{\zeta}_0 + \hat{\zeta}_1 l_i + \hat{\zeta}_2 l_i^2\right) = \exp((l_i - \hat{\mu}_{L2}) \times \hat{A1} + \left(l_i^2 - \hat{\mu}_{L2}^2 - \hat{\sigma}_{L2}^2\right) \times \hat{A2}),$$

for $i$ in $1,\ldots,N_l$.

Fitting an unadjusted logistic regression outcome model (i.e., $Y \sim 1$) to the study 1 IPD with observations weighted by the MAIC weights allows one to obtain $\hat{\theta}_{AC|Study2,MAIC}$. The estimated marginal treatment effect of $C$ vs. $B$ in the study 2 population is then calculated as: $\hat{\theta}_{CB|Study2,MAIC} = \hat{\theta}_{AC|Study2,MAIC} - \hat{\theta}_{AB|Study2}$. For further details see Remiro-Azócar (2022) and Appendix D of Phillippo et al. (2016).

The **STC** approach for binary outcome data involves three steps. The first step is to fit a logistic regression outcome model with the study 1 data conditional on $L$ (either with or without a treatment-covariate interaction term) to obtain estimates of the conditional probabilities: $\hat{\Pr}(Y = 1|trt = A, L = l)$, and $\hat{\Pr}(Y = 1|trt = C, L = l)$, for any arbitrary value of $l$. The second step is to simulate a large number, $W$, of pseudo covariate values, $l^*$, from a distribution that matches as best as possible the distribution of the $L$ values observed in study 2. For Simulation Study 2, we simulate $W$=10,000 pseudo-values from a normal distribution with mean and standard deviation equal to $\hat{\mu}_{L2}$ and $\hat{\sigma}_{L2}$, respectively. Recall again that $\hat{\mu}_{L2}$ and $\hat{\sigma}_{L2}$ are available from the study 2 AD. Then, in the third step, one averages over these pseudo-values to obtain estimates of the marginal probabilities of $Y$ given treatment $A$ and given treatment $C$ under the study 2 population:

$$\hat{\Pr}(Y = 1|trt = A)_{Study2,STC} = \frac{1}{W} \sum_{i=1}^{W} \hat{\Pr}(Y = 1|trt = A, L = l_i^*),$$

and

$$\hat{\Pr}(Y = 1|trt = C)_{Study2,STC} = \frac{1}{W} \sum_{i=1}^{W} \hat{\Pr}(Y = 1|trt = C, L = l_i^*).$$

These can be used to calculate the covariate-adjusted estimator of the marginal log-OR for treatment $A$ vs. $C$ under the study 2 population:

$$\hat{\theta}_{AC|Study2,STC} = \log\left(\frac{\hat{\Pr}(Y = 1|\text{trt} = A)_{Study2,STC}}{1 - \hat{\Pr}(Y = 1|\text{trt} = A)_{Study2,STC}}\right) - \log\left(\frac{\hat{\Pr}(Y = 1|\text{trt} = C)_{Study2,STC}}{1 - \hat{\Pr}(Y = 1|\text{trt} = C)_{Study2,STC}}\right).$$

Finally, the estimated marginal treatment effect of *C* vs. *B* in the study 2 population is calculated as: $\hat{\theta}_{CB|Study2,STC} = \hat{\theta}_{AC|Study2,STC} - \hat{\theta}_{AB|Study2}$. Note that the implicitly defined propensity score model in STC assumes trial selection is not dependent on any covariates (i.e., *Study* ~ 1).

For TTE outcome data, the STC procedure is similar. The first two steps are identical, the only difference is that a Cox regression is fit in step 1 as the outcome model instead of a logistic regression. To average over the pseudo-values in the third step, one must use non-parametric Monte Carlo integration following the fourteen-step procedure proposed by Daniel et al. (2021). As noted earlier, this approach (conducting STC with Daniel et al. (2021)'s standardization to obtain a marginal estimate) is rather novel and has yet to be applied in any published ITCs.

### 3.5 Sample size, simulations, and performance measures

In each scenario, the study sample size is set to *N*=1000 for both study 1 and study 2. The covariate values are simulated anew for each simulation run (this follows what is done in previous MAIC vs. STC simulation studies; e.g., Remiro-Azócar et al. (2022)). For each of the 16 configurations (8 scenarios, 2 outcome types (binary and TTE)), we simulated 10000 unique datasets and calculated the marginal relative treatment effect estimate (for *B* vs. *C* in the study 2 population) using each of the 5 estimation approaches (A1-A5). We summarize results by calculating, for each of the 5 estimation approaches, the sample mean and standard deviation of these 10000 estimates, which respectively are our simulation estimators of the mean and empirical standard error of each estimator. We also calculate the Monte Carlo SE of both simulation estimators using the formulae given in Morris et al. (2019). We quote our results to 3 decimal places.

### 3.6 Results

Results for the TTE outcome are listed in Table 2. Results for the binary outcome are all very similar and are listed in Table A2 in the Appendix.

First, the results of Simulation Study 2 for TTE outcome data suggest that one can use the standardization method of Daniel et al. (2021) as an extension to conventional STC for the estimation of marginal HRs in an anchored ITC. The results suggest that in an STC, the standardization method of Daniel et al. (2021) works for TTE outcome data just as effectively as the standardization method proposed by Remiro-Azócar (2022) ("g-computation") for binary outcome data.

With regards to bias, we see the very same pattern in terms of which methods are and are not biased that was saw in Simulation Study 1. All five of the approaches appear to be unbiased if and only if *either* the propensity score model *or* the outcome model is correctly specified. However, the magnitude of bias is, in many instances, much smaller than what was observed in Simulation Study 1. Consider Scenario SS-2A in which *L* is a prognostic factor that is differentially distributed across trials (different mean, but same variance). In this scenario, the Bucher approach (A1) is biased, but the magnitude of the bias is much less than what was observed with the univariate approach in Simulation Study 1. Indeed, in Simulation Study 1, the bias observed with the univariate approach was 1.313 (=2.081-0.768) and in Simulation Study 2 the bias observed with the Bucher approach was 0.074 (=0.842-0.768). This is because much of the bias observed in Simulation Study 1 was due to confounding. In Simulation Study 2, since both studies in the ITC are RCTs, confounding bias is no longer a factor. The relatively smaller bias that we observe in Simulation Study 2 is instead due to the non-collapsibility of the OR and HR.

Another way to think about this is that the Bucher method (A1) is biased because neither the outcome model (*Y~X*), nor the propensity model (*Study~1*) defined include a *L* term and are therefore *both* misspecified. Despite having misspecified outcome models, both MAIC methods (A2 and A3) define correctly specified propensity score models (*Study~L* and *Study~L+L$^2$*) and are therefore unbiased. Both standardization methods (A4 and A5) are also unbiased, despite having misspecified propensity score models, since the outcome models they define (*Y~X+L* and *Y~X+L+XL*) are correctly specified.

With regards to efficiency, there are also many similarities with what was observed in Simulation Study 1. When the Bucher approach (A1) is unbiased (in SS-1A and SS-1B), it is at least as efficient as all the other methods. This parallels how, in Simulation Study 1, the univariate approach, when unbiased is always at least as efficient as the PSW and standardization approaches. When both MAIC and STC approaches are unbiased, STC is always at least as efficient as MAIC. This corresponds to how, in Simulation Study 1, standardization is always at least as efficient as PSW, when both approaches are unbiased. Finally, when both STC approaches (A4 and A5) are unbiased (in SS-1A, SS-1B, SS-2A, and SS-2B), the STC without interaction approach (A4) is always at least as efficient as the STC with interaction (A5). This suggests that there is a cost to overfitting the outcome model in a STC. The very same result was observed for standardization in Simulation Study 1.

There is one notable difference between thew results of Simulation Study 1 and Simulation Study 2 with regards to efficiency. When both MAIC approaches (A2 and A3) are unbiased (in SS-1A, SS-1B, SS-2A, SS-3A and SS-4A) the MAIC matching on only 1$^{st}$ moment (A2) is always more efficient than the MAIC matching on both 1$^{st}$ and 2$^{nd}$ moments (A3). This suggests that there is indeed a cost to unnecessarily matching on both 1$^{st}$ and 2$^{nd}$ moments (i.e., a cost to overfitting the trial allocation model). This is contrary to what was observed for PSW in Simulation Study 1.

## 4 Discussion

### 4.1 Marginalization in the context of a single comparative study

The results of Simulation Study 1 and Simulation Study 2 indicate that many of the principles that apply to estimating marginal relative treatment effects from a single observational study also apply to estimating marginal relative treatment effects from an anchored indirect treatment comparison. This highlights the notion that ITCs are "essentially observational findings across trials" (Higgins and Green, 2011). Specifically, we found that all of the approaches we considered will be unbiased if and only if *either* the propensity score model *or* the outcome model is correctly specified. For MAIC specifically, Cheng et al. (2022) arrive at the same conclusion (MAIC is unbiased "when either the trial selection model or an implicit linear mean outcome model is correct") and therefore claim that "MAIC is thus more robust than STC." In our assessment, since STC is also unbiased when either the trial selection model or the outcome model is correct, we do not see either method as more or less robust than the other. Rather, each is robust to a different kind of model misspecification. Unless there are no unbalanced prognostic variables and no unbalanced effect modifiers, MAIC will be biased whenever the trial allocation model is misspecified, whereas STC will be biased whenever the outcome model is misspecified. Our results also suggest that, in scenarios in which both methods are unbiased, STC will be more efficient than MAIC.

| | Scenario | | True marginal logHR $\theta_{CB\|Study2}$ | Performance measure | A1. Bucher OM: $Y\sim X$ PS: $Sdy\sim 1$ | A2. MAIC (1st moment) OM: $Y\sim X$ PS: $Sdy\sim L$ | A3. MAIC (1st + 2nd) OM: $Y\sim X$ PS: $Sdy\sim L+L^2$ | A4. STC OM: $Y\sim X+L$ PS: $Sdy\sim 1$ | A5. STC with interaction OM: $Y\sim X+L+XL$ PS: $Sdy\sim 1$ |
|---|---|---|---|---|---|---|---|---|---|
| **Indirect Treatment Comparison** | **ITC-1A.** OM: $Y\sim X$ PS: $Study\sim L$ | $\beta_1=0$ $\beta_{2AC}=0$ $\beta_{3AC}=0$ $\kappa_1=1$ $\kappa_2=0$ | 1.000 | Mean (MC error) | 1.001 (0.0009) | 1.007 (0.0016) | 1.007 (0.0016) | 1.001 (0.0009) | 1.001 (0.0013) |
| | | | | Empirical SE (MC error) | 0.095 (0.0007) | 0.160 (0.0011) | 0.165 (0.0012) | 0.095 (0.0007) | 0.126 (0.0009) |
| | **ITC-1B.** OM: $Y\sim X$ PS: $Study\sim L^2$ | $\beta_1=0$ $\beta_{2AC}=0$ $\beta_{3AC}=0$ $\kappa_1=0$ $\kappa_2=1$ | 1.000 | Mean (MC error) | 1.000 (0.0010) | 1.000 (0.0010) | 1.003 (0.0012) | 1.000 (0.0010) | 1.001 (0.0010) |
| | | | | Empirical SE (MC error) | 0.096 (0.0007) | 0.096 (0.0007) | 0.120 (0.0008) | 0.096 (0.0007) | 0.096 (0.0007) |
| | **ITC-2A.** OM: $Y\sim X+L$ PS: $Study\sim L$ | $\beta_1=1$ $\beta_{2AC}=0$ $\beta_{3AC}=0$ $\kappa_1=1$ $\kappa_2=0$ | 0.612 | Mean (MC error) | <span style="color:red">0.566 (0.0010)</span> | 0.625 (0.0023) | 0.618 (0.0023) | 0.613 (0.0009) | 0.613 (0.0011) |
| | | | | Empirical SE (MC error) | <span style="color:red">0.103 (0.0007)</span> | 0.227 (0.0016) | 0.233 (0.0016) | 0.090 (0.0006) | 0.111 (0.0008) |
| | **ITC-2B.** OM: $Y\sim X+L$ PS: $Study\sim L^2$ | $\beta_1=1$ $\beta_{2AC}=0$ $\beta_{3AC}=0$ $\kappa_1=0$ $\kappa_2=1$ | 0.759 | Mean (MC error) | <span style="color:red">0.486 (0.0010)</span> | <span style="color:red">0.486 (0.0010)</span> | 0.761 (0.0011) | 0.761 (0.0009) | 0.761 (0.0009) |
| | | | | Empirical SE (MC error) | 0.098 (0.0007) | 0.098 (0.0007) | 0.110 (0.0008) | 0.087 (0.0006) | 0.090 (0.0006) |
| | **ITC-3A.** OM: $Y\sim X+L+XL$ PS: $Study\sim L$ | $\beta_1=1$ $\beta_{2AC}=1$ $\beta_{3AC}=0$ $\kappa_1=1$ $\kappa_2=0$ | 0.079 | Empirical SE (MC error) | <span style="color:red">0.503 (0.0010)</span> | 0.080 (0.0022) | 0.072 (0.0022) | <span style="color:red">0.902 (0.0009)</span> | 0.074 (0.0010) |
| | | | | Empirical SE (MC error) | 0.104 (0.0007) | 0.215 (0.0015) | 0.224 (0.0016) | <span style="color:red">0.090 (0.0006)</span> | 0.103 (0.0007) |
| | **ITC-3B.** OM: $Y\sim X+L+XL$ PS: $Study\sim L^2$ | $\beta_1=1$ $\beta_{2AC}=1$ $\beta_{3AC}=0$ $\kappa_1=0$ $\kappa_2=1$ | 0.459 | Mean (MC error) | <span style="color:red">0.187 (0.0010)</span> | <span style="color:red">0.187 (0.0010)</span> | 0.461 (0.0011) | <span style="color:red">0.920 (0.0009)</span> | 0.467 (0.0009) |
| | | | | Empirical SE (MC error) | <span style="color:red">0.101 (0.0007)</span> | <span style="color:red">0.100 (0.0007)</span> | 0.109 (0.0008) | <span style="color:red">0.091 (0.0006)</span> | 0.090 (0.0006) |
| | **ITC-4A.** OM: $Y\sim X+L+XL^2$ PS: $Study\sim L$ | $\beta_1=1$ $\beta_{2AC}=0$ $\beta_{3AC}=1$ $\kappa_1=1$ $\kappa_2=0$ | 1.807 | Mean (MC error) | <span style="color:red">1.017 (0.0010)</span> | 1.816 (0.0014) | 1.829 (0.0015) | <span style="color:red">1.039 (0.0009)</span> | <span style="color:red">0.232 (0.0012)</span> |
| | | | | Empirical SE (MC error) | 0.103 (0.0007) | 0.145 (0.0010) | 0.148 (0.0010) | <span style="color:red">0.090 (0.0006)</span> | <span style="color:red">0.121 (0.0009)</span> |
| | **ITC-4B.** OM: $Y\sim X+L+XL^2$ PS: $Study\sim L^2$ | $\beta_1=1$ $\beta_{2AC}=0$ $\beta_{3AC}=1$ $\kappa_1=0$ $\kappa_2=1$ | 1.140 | Mean (MC error) | <span style="color:red">1.445 (0.0010)</span> | <span style="color:red">1.444 (0.0009)</span> | 1.144 (0.0012) | <span style="color:red">1.448 (0.0009)</span> | <span style="color:red">1.581 (0.0011)</span> |
| | | | | Empirical SE (MC error) | <span style="color:red">0.098 (0.0007)</span> | <span style="color:red">0.095 (0.0007)</span> | 0.123 (0.0009) | <span style="color:red">0.094 (0.0007)</span> | <span style="color:red">0.108 (0.0008)</span> |

**Table 2** – Results from Simulation Study 2 for TTE outcome data; results obtained for the marginal log-HR. Values listed under "True marginal logHR, $\theta_{CB|Study2}$" were calculated by approximation by simulating 10 million observations from the true Weibull sampling distributions. Numbers in red correspond to results for which the difference between the mean and the true marginal log-OR is more than 0.1.

> **Key Findings**
>
> - The standardization method of Daniel et al. (2021) can be used as an extension to conventional STC for the estimation of marginal hazard ratios in an anchored ITC.
>
> - To obtain unbiased estimates of the marginal odds ratio or hazard ratio in an anchored ITC, one must adjust for all unbalanced prognostic variables in addition to all unbalanced effect modifiers.
>
> - Unless there are no unbalanced prognostic variables and no unbalanced effect modifiers, MAIC will be biased whenever the trial allocation model is misspecified. It may therefore be necessary to match on both the 1$^{st}$ and 2$^{nd}$ moments in a MAIC.
>
> - Unless there are no unbalanced prognostic variables and no unbalanced effect modifiers, STC will be biased whenever the outcome model is misspecified. It may therefore be necessary to adjust for interactions and/or higher order terms.

Despite not needing to worry about confounding bias in an ITC of RCTs, bias due to non-collapsibility may be an issue. Specifically, the results of Simulation Study 2 suggest that to obtain unbiased OR and HR estimates in an anchored ITC, one must adjust for *all* unbalanced prognostic variables (in addition to *all* unbalanced effect modifiers). This is contrary to the recommendations for using MAIC of Vo (2023) (purely prognostic variables can be "safely excluded") and NICE DSU guidelines ("To avoid loss of precision due to over-matching, no prognostic variables which are not also effect modifiers should be adjusted for, as variables which are purely prognostic do not affect the estimated relative treatment effect." (Phillipo et al., 2016)). Whether or not adjusting for *all* unbalanced prognostic variables and *all* effect modifiers is realistic will no doubt depend on the specific context. Researchers may have been under the impression that such a strong requirement was only necessary for *unanchored* ITCs (Phillipo et al. (2016): "in all cases where unanchored indirect comparisons are performed, a strong assumption is made that all prognostic variables and all effect modifiers are accounted for and correctly specified – an assumption largely considered to be implausibly strong."). Our results suggest that this "strong assumption" is also needed in anchored ITCs for any non-collapsible estimands. Recently, there has been increased attention on the importance of correctly identifying effect modifiers (e.g., Frietag et al., 2023; Nguyen et al., 2018). Greater attention should also be focussed on correctly identifying purely prognostic variables.

Finally, depending on the nature of the true trial allocation model, the results of Simulation Study 2 suggest that it may be necessary to match on both the 1$^{st}$ and 2$^{nd}$ moments in a MAIC. This is contrary to some current recommendations (e.g., the conclusions of Hatswell et al., 2020). Unfortunately, matching on both 1$^{st}$ and 2$^{nd}$ moments when matching on only the 1$^{st}$ moment is sufficient may be costly in terms of efficiency. This is contrary to what is known (and what we observed in Simulation Study 1) about the consequences of overfitting one's propensity score model in the context of a single observational study. This difference may be a result of using the method of moments to calculate weights for the MAIC.

### 4.3 Limitations and recommendations

Our simulation studies have a few notable limitations. First, certain design choices in the simulation study are perhaps unrealistic. Specifically, the follow-up times and study sample sizes are likely overly optimistic; see Tai et al. (2021). As in Daniel et al. (2021)'s simulation study, we selected large sample sizes to ensure that finite sample bias (Greenland, Manrournia, & Altman, 2016) would be negligible. A follow-up simulation study to determine how the methods compare with small sample sizes would be of interest. Our simulation setup was also rather unrealistic with respect to the fact that we only considered a single covariate. This was a deliberate choice to make the results as straightforward as possible. We note that Chatton et al. (2020) who compared different methods for estimating the marginal OR in a simulation study with nine covariates (but without any effect modifiers) arrived at some similar conclusions: standardization (with a correctly specified outcome model) is the method with lowest bias and lowest variance. More recently, Chatton et al. (2022) considered data with time-to-event outcomes. Based on a simulation study with six covariates (but without any effect modifiers), they concluded that standardization (with a correctly specified outcome model) is the most efficient approach for estimating the restricted mean survival time (RSMT) and the so-called "average hazard ratio" (AHR) (two alternatives to the marginal HR).

A second limitation of our simulation studies is that, while we considered the biasedness and efficiency of point estimates, we did not measure the coverage and width of confidence intervals. The non-parametric bootstrap can be used to calculate confidence intervals for both the PSW approaches (Austin, 2016; Remiro-Azócar et al., 2021; Matsouaka et al., 2023) and standardization-based approaches (Daniel et al, 2021; Remiro-Azócar et al., 2022). While there are other, non-bootstrap approaches for calculating confidence intervals when using PSW-based approaches (e.g., using a model-based variance estimator, or using a robust sandwich-type variance estimator), these alternative approaches may lead to incorrect coverage rates (Austin, 2016; Remiro-Azócar et al., 2021). We also note that, when bootstrapping for STC, researchers must be careful to adequately take into account the uncertainty involved in the pseudo-data generation. A simulation study to evaluate the performance of bootstrap-based confidence intervals in the different settings we considered is an objective for future research (and will notably require substantial computational resources).

A third limitation is that we only considered the scenario of an anchored ITC and did not investigate what might occur with an unanchored ITC when there is no common comparator arm, or when an external control arm is used to compare against a single-arm trial (Ren et al., 2023). That being said, an unanchored ITC is not unlike a two-arm observational study (which we did consider in Simulation Study 1) where the probability of receiving a specific treatment (or of being in a specific study) may be confounded. Further research should investigate to what extent our findings can be generalized more broadly, for instance to large (network) meta-analyses, where there are multiple studies of which some are IPD; see Phillippo et al. (2020), Vo et al. (2019), and Schnitzer et al. (2016).

Finally, we focussed only on the most well-known PSW and standardization-based methods and there are other approaches that we did not consider. These include the recently proposed two-stage MAIC (Remiro-Azócar, 2022), entropy-based methods (Josey et al., 2021), and targeted maximum likelihood estimation (TMLE) (Smith et al., 2023, Schuler and Rose, 2017). Tackney et al. (2023) conduct a simulation study to compare a number of different approaches under model misspecification in RCTs with small sample sizes. Further research is needed to better understand how these methods compare more broadly.

Our overall conclusion is that estimating marginal treatment effects when the parameter of interest is non-collapsible can be challenging. While certain statistical methods can be used to obtain unbiased marginal estimates in certain scenarios, researchers must properly understand their many limitations. PSW and MAIC approaches are becoming increasingly popular. Webster-Clark et al. (2021) report an approximately 28-fold increase in the number of papers applying PSW methods between 2004 and 2018. By contrast, using standardization-based approaches is relatively rare for single comparator studies (Snowden et al., 2011), and the concept of implementing STC along with standardization in order to obtain a marginal effect estimate has only recently been proposed (Remiro-Azócar et al., 2022). Indeed, Daniel et al. (2021) describe standardization as a "largely ignored procedure." This strikes us as somewhat unfortunate, given the results of our simulation studies and an earlier understanding that, in general, regression adjustment methods can often be more efficient than methods based on weighting (Robins et al., 1992). Furthermore, given the increasing interest in doubly-robust estimators for single observational studies (Funk et al., 2011), it is rather surprising that doubly-robust estimators have not yet been proposed for ITCs.

# Appendix

| | Scenario | | True marginal log-OR | Performance measure | A1. Univ. OM: $Y\sim X$ PS: $X\sim 1$ | A2. PSW OM: $Y\sim X$ PS: $X\sim L$ | A3. PSW OM: $Y\sim X$ PS: $X\sim L+L^2$ | A4. std. OM: $Y\sim X+L$ PS: $X\sim 1$ | A5. std. OM: $Y\sim X+L+XL$ PS: $X\sim 1$ |
|---|---|---|---|---|---|---|---|---|---|
| **Observational Study** | **SS-1A.** OM: $Y\sim X$ PS: $X\sim L$ | $\beta_1=0$ $\beta_2=0$ $\beta_3=0$ $\kappa_1=1$ $\kappa_2=0$ | 1.000 | Mean (MC error) Empirical SE (MC error) | 1.004 (0.0012) 0.121 (0.0009) | 1.017 (0.0023) 0.229 (0.0016) | 1.017 (0.0023) 0.228 (0.0016) | 1.003 (0.0014) 0.142 (0.0010) | 1.005 (0.0017) 0.173 (0.0012) |
| | **SS-1B.** OM: $Y\sim X$ PS: $X\sim L^2$ | $\beta_1=0$ $\beta_2=0$ $\beta_3=0$ $\kappa_1=0$ $\kappa_2=1$ | 1.000 | Mean (MC error) Empirical SE (MC error) | 0.999 (0.0013) 0.129 (0.0009) | 0.999 (0.0013) 0.129 (0.0009) | 1.002 (0.0015) 0.153 (0.0011) | 1.001 (0.0013) 0.130 (0.0009) | 1.001 (0.0013) 0.130 (0.0009) |
| | **SS-2A.** OM: $Y\sim X+L$ PS: $X\sim L$ | $\beta_1=1$ $\beta_2=0$ $\beta_3=0$ $\kappa_1=1$ $\kappa_2=0$ | 0.768 | Mean (MC error) Empirical SE (MC error) | <span style="color:red">2.081 (0.0013) 0.126 (0.0009)</span> | 0.781 (0.0021) 0.207 (0.0015) | 0.786 (0.0019) 0.193 (0.0014) | 0.770 (0.0012) 0.118 (0.0008) | 0.772 (0.0014) 0.141 (0.0010) |
| | **SS-2B.** OM: $Y\sim X+L$ PS: $X\sim L^2$ | $\beta_1=1$ $\beta_2=0$ $\beta_3=0$ $\kappa_1=0$ $\kappa_2=1$ | 0.925 | Mean (MC error) Empirical SE (MC error) | <span style="color:red">0.444 (0.0012) 0.116 (0.0008)</span> | <span style="color:red">0.435 (0.0011) 0.110 (0.0008)</span> | 0.927 (0.0014) 0.139 (0.0010) | 0.926 (0.0013) 0.126 (0.0009) | 0.927 (0.0013) 0.127 (0.0009) |
| | **SS-3A.** OM: $Y\sim X+L+XL$ PS: $X\sim L$ | $\beta_1=1$ $\beta_2=1$ $\beta_3=0$ $\kappa_1=1$ $\kappa_2=0$ | 0.066 | Empirical SE (MC error) Empirical SE (MC error) | <span style="color:red">1.827 (0.0011) 0.115 (0.0008)</span> | 0.071 (0.0015) 0.150 (0.0011) | 0.074 (0.0012) 0.124 (0.0009) | <span style="color:red">0.368 (0.0010) 0.101 (0.0007)</span> | 0.066 (0.0010) 0.096 (0.0007) |
| | **SS-3B.** OM: $Y\sim X+L+XL$ PS: $X\sim L^2$ | $\beta_1=1$ $\beta_2=1$ $\beta_3=0$ $\kappa_1=0$ $\kappa_2=1$ | 0.575 | Mean (MC error) Empirical SE (MC error) | <span style="color:red">-0.118 (0.0011) 0.112 (0.0008)</span> | <span style="color:red">-0.129 (0.0010) 0.103 (0.0007)</span> | 0.576 (0.0013) 0.128 (0.0009) | <span style="color:red">0.483 (0.0012) 0.117 (0.0008)</span> | 0.576 (0.0012) 0.124 (0.0009) |
| | **SS-4A.** OM: $Y\sim X+L+XL^2$ PS: $X\sim L$ | $\beta_1=1$ $\beta_2=0$ $\beta_3=1$ $\kappa_1=1$ $\kappa_2=0$ | 2.360 | Mean (MC error) Empirical SE (MC error) | <span style="color:red">2.773 (0.0016) 0.160 (0.0011)</span> | 2.377 (0.0020) 0.200 (0.0014) | 2.373 (0.0020) 0.196 (0.0014) | <span style="color:red">1.468 (0.0016) 0.156 (0.0011)</span> | <span style="color:red">1.751 (0.0019) 0.194 (0.0014)</span> |
| | **SS-4B.** OM: $Y\sim X+L+XL^2$ PS: $X\sim L^2$ | $\beta_1=1$ $\beta_2=0$ $\beta_3=1$ $\kappa_1=0$ $\kappa_2=1$ | 1.356 | Mean (MC error) Empirical SE (MC error) | <span style="color:red">2.012 (0.0016) 0.156 (0.0011)</span> | <span style="color:red">2.010 (0.0015) 0.154 (0.0011)</span> | 1.364 (0.0016) 0.165 (0.0012) | <span style="color:red">2.131 (0.0016) 0.165 (0.0012)</span> | <span style="color:red">2.049 (0.0015) 0.155 (0.0011)</span> |

**Table A1** – Results from Simulation Study 1 for binary outcome data; results obtained for the marginal log-OR. Values listed under "True marginal log-OR" were calculated by approximation by simulating 10 million observations from the true Binomial sampling distributions. Numbers in red correspond to results for which the difference between the Mean and the true marginal log-OR is more than 0.1.

| | Scenario | | True marginal logOR $\theta_{CB|Study2}$ | Performance measure | A1. Bucher OM: $Y\sim X$ PS: $Sdy \sim 1$ | A2. MAIC (1st moment) OM: $Y\sim X$ PS: $Sdy \sim L$ | A3. MAIC (1st + 2nd) OM: $Y\sim X$ PS: $Sdy \sim L+L^2$ | A4. STC OM: $Y\sim X+L$ PS: $Sdy \sim 1$ | A5. STC with interaction OM: $Y\sim X+L+XL$ PS: $Sdy \sim 1$ |
|---|---|---|---|---|---|---|---|---|---|
| **Indirect Treatment Comparison** | **ITC-1A.** OM: $Y \sim X$ PS: $Sdy \sim L$ | $\beta_1 = 0$ $\beta_{2AC} = 0$ $\beta_{3AC} = 0$ $\kappa_1 = 1$ $\kappa_2 = 0$ | 1.000 | Mean (MC error) | 1.008 (0.0073) | 1.022 (0.0122) | 1.025 (0.0126) | 1.008 (0.0073) | 1.004 (0.0096) |
| | | | | Empirical SE (MC error) | 0.230 (0.0051) | 0.387 (0.0086) | 0.400 (0.0089) | 0.230 (0.0051) | 0.305 (0.0068) |
| | **ITC-1B.** OM: $Y \sim X$ PS: $Sdy \sim L^2$ | $\beta_1 = 0$ $\beta_{2AC} = 0$ $\beta_{3AC} = 0$ $\kappa_1 = 0$ $\kappa_2 = 1$ | 1.000 | Mean (MC error) | 1.010 (0.0068) | 1.010 (0.0068) | 1.014 (0.0093) | 1.011 (0.0068) | 1.014 (0.0069) |
| | | | | Empirical SE (MC error) | 0.216 (0.0048) | 0.216 (0.0048) | 0.293 (0.0066) | 0.216 (0.0048) | 0.216 (0.0048) |
| | **ITC-2A.** OM: $Y \sim X+L$ PS: $Sdy \sim L$ | $\beta_1 = 1$ $\beta_{2AC} = 0$ $\beta_{3AC} = 0$ $\kappa_1 = 1$ $\kappa_2 = 0$ | 0.768 | Mean (MC error) | <span style="color:red">0.842 (0.0074)</span> | 0.775 (0.0115) | 0.768 (0.0118) | 0.772 (0.0067) | 0.760 (0.0083) |
| | | | | Empirical SE (MC error) | <span style="color:red">0.234 (0.0052)</span> | 0.365 (0.0082) | 0.375 (0.0084) | 0.211 (0.0047) | 0.262 (0.0059) |
| | **ITC-2B.** OM: $Y \sim X+L$ PS: $Sdy \sim L^2$ | $\beta_1 = 1$ $\beta_{2AC} = 0$ $\beta_{3AC} = 0$ $\kappa_1 = 0$ $\kappa_2 = 1$ | 0.925 | Mean (MC error) | <span style="color:red">0.695 (0.0064)</span> | <span style="color:red">0.695 (0.0064)</span> | 0.932 (0.0087) | 0.929 (0.0069) | 0.932 (0.0075) |
| | | | | Empirical SE (MC error) | <span style="color:red">0.203 (0.0045)</span> | <span style="color:red">0.203 (0.0045)</span> | 0.274 (0.0061) | 0.219 (0.0049) | 0.236 (0.0053) |
| | **ITC-3A.** OM: $Y \sim X+L+XL$ PS: $Sdy \sim L$ | $\beta_1 = 1$ $\beta_{2AC} = 1$ $\beta_{3AC} = 0$ $\kappa_1 = 1$ $\kappa_2 = 0$ | 0.066 | Empirical SE (MC error) | <span style="color:red">0.594 (0.0072)</span> | 0.065 (0.0105) | 0.053 (0.0108) | <span style="color:red">0.533 (0.0061)</span> | 0.059 (0.0071) |
| | | | | Empirical SE (MC error) | <span style="color:red">0.227 (0.0051)</span> | 0.331 (0.0074) | 0.341 (0.0076) | <span style="color:red">0.194 (0.0043)</span> | 0.224 (0.0050) |
| | **ITC-3B.** OM: $Y \sim X+L+XL$ PS: $Sdy \sim L^2$ | $\beta_1 = 1$ $\beta_{2AC} = 1$ $\beta_{3AC} = 0$ $\kappa_1 = 0$ $\kappa_2 = 1$ | 0.575 | Mean (MC error) | <span style="color:red">0.131 (0.0061)</span> | <span style="color:red">0.131 (0.0061)</span> | 0.578 (0.0080) | <span style="color:red">0.196 (0.0066)</span> | 0.588 (0.0072) |
| | | | | Empirical SE (MC error) | <span style="color:red">0.193 (0.0043)</span> | <span style="color:red">0.193 (0.0043)</span> | 0.252 (0.0056) | <span style="color:red">0.209 (0.0047)</span> | 0.227 (0.0051) |
| | **ITC-4A.** OM: $Y \sim X+L+XL^2$ PS: $Sdy \sim L$ | $\beta_1 = 1$ $\beta_{2AC} = 0$ $\beta_{3AC} = 1$ $\kappa_1 = 1$ $\kappa_2 = 0$ | 2.360 | Mean (MC error) | <span style="color:red">1.529 (0.0087)</span> | 2.365 (0.0106) | 2.403 (0.0110) | <span style="color:red">1.412 (0.0080)</span> | <span style="color:red">1.746 (0.0101)</span> |
| | | | | Empirical SE (MC error) | <span style="color:red">0.275 (0.0061)</span> | 0.335 (0.0075) | 0.347 (0.0077) | <span style="color:red">0.254 (0.0057)</span> | <span style="color:red">0.320 (0.0072)</span> |
| | **ITC-4B.** OM: $Y \sim X+L+XL^2$ PS: $Sdy \sim L^2$ | $\beta_1 = 1$ $\beta_{2AC} = 0$ $\beta_{3AC} = 1$ $\kappa_1 = 0$ $\kappa_2 = 1$ | 1.356 | Mean (MC error) | <span style="color:red">2.264 (0.0088)</span> | <span style="color:red">2.263 (0.0088)</span> | 1.371 (0.0103) | <span style="color:red">2.626 (0.0102)</span> | <span style="color:red">2.043 (0.0090)</span> |
| | | | | Empirical SE (MC error) | <span style="color:red">0.279 (0.0062)</span> | <span style="color:red">0.278 (0.0062)</span> | 0.326 (0.0073) | <span style="color:red">0.322 (0.0072)</span> | <span style="color:red">0.285 (0.0064)</span> |

**Table A2** – Results from Simulation Study 2 for binary outcome data; results obtained for the marginal log-OR. Values listed under "True marginal logOR $\theta_{CB|Study2}$" were calculated by approximation by simulating 10 million observations from the true Binomial sampling distributions. Numbers in red correspond to results for which the difference between the Mean and the true marginal log-OR is more than 0.1.